# Molecular field approximation in theory of ferromagnetic phase transition in diluted magnetic semiconductors

Y. G. Semenov[1] and S. M. Ryabchenko[2]

[1]V. Lashkaryov Institute of Semiconductor Physics of National Academy of Sciences of Ukraine, Kyiv 03680, Ukraine

[2]Institute of Physics, of National Academy of Sciences of Ukraine, Kyiv 03028, Ukraine

## Abstract

In this pedagogical paper, the comparative analysis of two common approaches describing the ferromagnetic phase transition in diluted magnetic semiconductors (DMS) is expounded in terms of Weiss field approximation. Assuming a finite spin polarization of the magnetic ions, the treatment of carrier-ion exchange interaction in first order evokes the homogeneous Weiss molecular field that polarizes the spins of free carriers. In turn, this spin polarization of the free carriers exerts the effective field that may stabilize DMS spin polarization below a critical temperature $T_C$. The treatment of such self-consistent spontaneous DMS magnetization can be done in terms of spin-spin interaction independent on inter-ion distance and infinitesimal in thermodynamic limit. On the other hand, the taking additionally into account the second order effects of carrier-ion exchange interaction treat a Weiss field in term of the Ruderman–Kittel–Kasuya–Yosida indirect spin-spin interaction, which oscillates and not disappears at finite inter-ion distance in case of finite carriers concentration. These both approaches result in same Curie temperature $T_C$ provided non-correlated homogeneous random distribution of the localized spin moments over sample volume. We discuss the origin of such coincidence and show when this is not a case in other more realistic models of the conducting DMSs.

## 1. Introduction

The Carrier-Ion Exchange Interaction (CIEI) triggered a train of unusual electronic phenomena in diluted magnetic semiconductors (DMS) like the $A_{1-x}Mn_xB$- compounds where AB denotes II – VI or III – V semiconductors [1, 2]. Shortly after discover the giant spin splitting of the exciton spectra in pioneer paper [1], the ferromagnetic (FM) phase transition accompanied by mutual spin polarization of band carriers and localized spin moments (LSMs) in DMS was predicted [3]. That paper had shown how to convert the CIEI of spin polarized LSM to the Weiss molecular field, which polarizes the spins of free carriers (electrons or holes). They, in turn, serve as a source of the exchange field capable to stabilize a finite magnetic moment of LSMs at certain temperature. The approach of Ref. [3] predicts the FM phase transition with simultaneous arising of the spontaneous magnetization for subsystems both the LSMs and the carriers (i.e. electrons and/or holes). Their mutual directions of spin polarizations depend on sign of CIEI that is generally different for electrons and holes.

In similar to papers [4,5] where modeling of the carriers-induced ferromagnetism through the CIEI was developed for the first time to describe the ferromagnetism of transition metals, the formalism of Ref. [3] involves an additional effect of ion-ion spin-spin interaction with exchange constants $J_{mm}(\mathbf{R}_{j1},\mathbf{R}_{j2})$, where $\mathbf{R}_{j1},\mathbf{R}_{j2}$ are the radii-vectors of the corresponding magnetic ions. In Ref. [3] one was expressed supposition that effective indirect RKKY [6–8] ion-ion exchange interaction through the carriers could be added to contributions of $J_{mm}(\mathbf{R}_{j1},\mathbf{R}_{j2})$. This not entirely correct assumption was not accompanied by more thorough theoretical analysis of [8, 9] as well as discussion presented in the rest of this paper. Moreover, all estimations in [3] were carried out in the limit $J_{mm} \to 0$. Consequently, the approach of Ref. [3] allows correctly estimate the range of Curie temperature Tc for actual magnetically doped semiconductors in the first order perturbation only.

The formalism developed in both Refs. [4,5] and Ref. [3] treats the CIEI in the first order of perturbation, which evaluates only corrections to the energy of interacted particles while it cannot account the effects of electron (hole) scattering with changing of their wave vectors. That is why a number of authors use the lexis "direct carrier-ions exchange interaction" to indicate CIEI in first order of perturbation. In this approach, the Weiss field generated by one LSM via free carriers affects another spins regardless the inter-ions distance. At one time, this circumstance caused a lively scientific discussion (see Refs [8] and [9]). Let us also mention that formalism described above does not involve modification of the LSM-LSM exchange interaction due to CIEI.

An alternative approach to this problem treats the free carriers as a mediator of indirect interaction between pairs of localized spins. Such interaction usually cites as Ruderman–Kittel–Kasuya–Yosida (RKKY) [6–8] indirect spin-spin interaction. The standard procedure presumes evaluating the energy of indirect interaction in the manner of second order of perturbation theory that mixes the electronic states with different wave vectors $\mathbf{k} \neq \mathbf{k}'$ but excludes the case $\mathbf{k}=\mathbf{k}'$. Accordingly, the spin-dependent second order correction to the energy oscillates with distance between spin locations. Note however, that actual RKKY-procedure also implicitly involves the contribution of singular points $\mathbf{k}=\mathbf{k}'$ of scattering without wave vector changing. That contribution was shown to coincide with first order effects of spin-spin interaction [8,9]. Thus, the RKKY indirect spin-spin interaction may lead to FM correlations and phase transition calculated in terms of first and second orders of perturbation theory.

The FM phase transitions in DMS were first observed in IV-Mn-VI compounds [10] and were explained in terms of the RKKY interactions. Surprisingly the Weiss molecular field calculated in terms of first order perturbation [3] and in terms of the sum of RKKY spin-spin interactions results in same expression in the case of random non-correlated distribution of the magnetic ions. Based on this particular statement, the identity of both approaches became a conventional wisdom [11]. In other words, the Curie temperature evaluated in first order of CIEI supposes to be not refined with adding the second order effects of RKKY interaction. This also means that second order effect represented by sum over scattering electrons/holes with $\mathbf{k} \neq \mathbf{k}'$ does not contribute to the energy of spins in the Weiss field under random non-correlated distribution of the magnetic ions.

Nevertheless, in contrast to common standpoint, the testing of this assumption for exactly solvable model of flat energy bands reveals the complementary contributions of the «direct » (1-th order) and «indirect» (2-th order) interactions to the expression of critical temperature of FM phase transition [12]. The intrigue of this antinomy was strengthened by the works [13, 14] demonstrated significant discrepancy of mean field [3] and RKKY treatment in more advanced random fields approximation. However the problem how approximation of band structure by more realistic than dispersionless one [12] would modify the Weiss molecular field still remains an actual problem of the theory of FM phase transition in disordered magnetic systems. Below we present the analysis of this problem in terms of the effective fields mediated by CIEI in first and second order separately. In particular, we show that second order correction to Weiss field becomes zero provided only non-correlated equiprobable LSM distribution up to infinitesimal inter-ion distances. Moreover, in contrast to such free gas modeling of LSM distribution the irremovable limitation of the minimal inter-ion distance by lattice constant may visibly correct the estimation of the critical temperature in DMS.

Note that giant spin splitting observed in DMS band structure is usually treated in first order on CIEI. This approach does not take into account the oscillations of spin density of the charge carrier around the LSM depending on the distance to it. On the other hand, such oscillations may be important in other phenomenon. For example, RKKY-like oscillations of the effective spin-spin interaction are responsible for non-monotonic dependence of the exchange interaction between FM layers on thickness of the non-magnetic spacer between them [15 – 18].

## 2. Weiss field in first order approximation

For definiteness sake, let us consider a DMS with degenerating gas of free carriers, electrons or holes, in a simple energy band accounted for an isotropy dispersion law with an effective mass $m^*$. Moreover, to focus only on the effects of CIEI, the direct spin-spin interactions not related to free carriers will drop out of consideration. This simplification merely corresponds to the model developed in Ref. [3] provided nulling the direct spin-spin interaction. The Hamiltonian of carrier-ion exchange interaction takes the form of contact interaction between electrons with spin-space coordinates $s_i, r_i$ and $N_m$ magnetic ions with spins $S_j$ situated at the lattice sites $R_j$:

$$\hat{H}_{L,e} = -\beta \sum_i \sum_j S_j s_i \delta(r_i - R_j). \tag{1}$$

The constant of CIEI $\beta N_0$, where $N_0$ is the primitive cells concentration (number of primitive cells divided on the volume) and $\beta$ is the calculation parameter with dimension of energy multiplied on volume, achieves a sufficient magnitude in the valence bands of most DMS up to 1 eV and even more [19]. The exchange integral, which follows from (1) depends on both crystal and magnetic ion parameters. The correspondent spin-Hamiltonian stems from the Eq. (1) after averaging the $\delta(r_i - R_j)$ on the carrier coordinate wave function $\Psi(r)$. The latter represents a standard Bloch function normalized on crystal volume $V$ if the CIEI (1) does not sufficiently disturb the electronic states. Necessary condition to this approximation corresponds to smallness of the parameter $\varepsilon = \beta N_0/W$, where $W$ is a relevant bandwidth [20]. Below this parameter assumes to be small and Bloch functions become good zero order approximation for the electron/hole states in DMS. One should note that exchange integral for one carrier with one LSM, $J_{1,1}$, is in such case

$$J_{1,1} = \beta/V. \tag{1a}$$

If the carrier is localized in part of the crystal, for instance the electron of neutral donor, or the carrier in case of magnetic polaron formation, the exchange integral of one carrier with one LSM may enhanced sufficiently due to growth of the carrier density on the LSM in decreased localization volume.

If one will neglect spin fluctuations of LSM, the Eq. (1) represents Zeeman energy of electrons in the homogeneous Weiss field of magnetic ions (m) acting on the carrier (let us it is electron (e)) spins:

$$B_{e/m} = \frac{\beta n_m \langle S_m \rangle}{g_e \mu_B}, \tag{2}$$

where $n_m$, $g_e$ and $\mu_B$ are the LSM concentration, electron g-factor and Bohr magneton respectively.

On the other hand, the CIEI (1) can be transform to Zeeman energy of the magnetic ions affected by the exchange field of the electrons with mean value $\langle s_e \rangle$ of spin polarization,

$$B_{m/e} = \frac{\beta n_e \langle s_e \rangle}{g_m \mu_B}, \tag{3}$$

where $g_m$ is LSM g-factor and $n_e$ is an electron concentration.

Equations (2) and (3) look like the complete system of the equations with respect to effective fields if one takes into account the relations

$$\langle S_m \rangle = -\chi_m^{(1)} \frac{B_{m/e}}{g_m \mu_B} \tag{4}$$

and

$$\langle S_e \rangle = -\chi_e^{(1)} \frac{B_{e/m}}{g_e \mu_B}, \qquad (5)$$

where LSM susceptibility $\chi_m^{(1)}(B_{m/e})$ per one LSM is a function of $B_{m/e}$ and temperature $T$ and $\chi_e^{(1)} = (3/8)g_e^2 \mu_B^2/\varepsilon_F$ is susceptibility per one particle of degenerated electrons with Fermi energy $\varepsilon_F$.

Equation for critical temperature $T_C$ of spontaneous magnetization appearance represents the case of infinitesimal exchange fields $B_{m/e}$ and $B_{e/m}$ in the Eqs (4) and (5). Thus, to find $T_C$ we should replace the $\chi_m^{(1)}(B_{m/h})$ by zero-field susceptibility $\chi_m^{(1)}(0)$. In the simplest case of magnetic ions interacted only via band carriers, solution of correspondent equation takes the form [12]:

$$T_c = \frac{S(S+1)}{3}\beta^2 n_m n_e \frac{\chi_e^{(1)}}{g_e^2 \mu_B^2} = \frac{S(S+1)}{8}\frac{\beta^2 n_m n_e}{\varepsilon_F}. \qquad (6)$$

Hereinafter, the temperature $T$ is expressed in energy units.

This equation can be derived in different manner useful for following analysis. Let us write down the Zeeman energy Hamiltonian $\hat{H}_{Zeem}^{eff} = g_m \mu_B B_{m/e} \sum_j S_j$ of the magnetic ions in the effective magnetic field [Eq. (3)] of conducting electrons. Then substitution of the Eq.(5) to Eq. (3) transforms Zeeman Hamiltonian to

$$\hat{H}_{Zeem}^{eff} = -\beta^2 n_m n_e \frac{\chi_e^{(1)}}{g_e^2 \mu_B^2} \langle S_m \rangle \sum_j S_j. \qquad (7)$$

Last equation clearly shows that Hamiltonian $\hat{H}_{Zeem}^{eff}$ is nothing but the Weiss-field approximation of the Hamiltonian of pair spin-spin interaction of $N_m$ LSMs diluted in a sample of volume $V$:

$$\hat{H}_{SS}^{(1)} = -\frac{1}{V}\frac{\beta^2 n_e \chi_e^{(1)}}{2 g_e^2 \mu_B^2} \sum_{j,j'(j \neq j')} S_j S_{j'}. \qquad (8)$$

These equations define the Weiss (or molecular) field

$$B_W^{(1)} = -\beta^2 \frac{n_m n_e}{g_m \mu_B} \frac{\chi_e^{(1)}}{g_e^2 \mu_B^2} \langle S_m \rangle \qquad (9)$$

that uniformly affects each LSM. Superscript "(1)" in the left sides of Eq. (8) and (9) indicates this result is valid in first order perturbation theory in terms of unperturbed electron eigenfunctions. The factor $1/2$ in Eq. (8) reflects the fact that each pair of spins in sum over the $j$ and $j'$ is twice accounted.

It would be instructive to show another derivation of the Eq. (8). It originates from definition of a homogeneous mean field induced by the ensemble of LSMs in operator representation $\hat{B}_{e/m} = \frac{\beta}{g_e \mu_B}\frac{1}{V}\sum_j S_j$. This field reduces the magnetic energy of the free carriers in standard form

$$\hat{H}_{SS}^{(1)} = -\frac{1}{2}V\chi_{Pauli}\hat{B}_{e/m}^2 \qquad (10)$$

that is just Eq. (8) provided the Pauli susceptibility of electron gas is $\chi_{Pauli} = n_e \chi_e^{(1)}$.

Eq. (10) establishes an effective spin-spin interaction independent on inter-ion distance. Moreover, it becomes infinitesimal in the thermodynamic limit $V \to \infty$ for any pair of spins that keeps a finite value of Weiss field in thermodynamic limit $N_m \to \infty$ under $n_m$=const, as mentioned above. This result arises in first order of perturbation theory that treats the CIEI (1) in terms of undisturbed electron wave functions. Nevertheless, a finite critical temperature (6) follows from the Weiss field approximation of the Eq.( 8) provided the finite LSM concentration $n_m = N_m/V$.

## 3. RKKY interaction

Alternative approach to FM phase transition stems from the CIEI transforming to indirect LSM spin-spin interaction mediated by free carriers in second order perturbation theory:

$$\hat{H}_{SS}^{(2)} = \frac{\beta^2}{V^2} \sum_{\sigma,\sigma'} \sum_{k,k'(k'\neq k)} \frac{f(\varepsilon_k)(1-f(\varepsilon_{k'}))}{\varepsilon_k - \varepsilon_{k'}} \sum_{j,j'(\neq j)} e^{i(k-k')(R_j - R_{j'})} (\sigma|S_j s_e|\sigma')(\sigma'|S_{j'} s_e|\sigma). \quad (11)$$

where $f(\varepsilon_k)$ is the Fermi-Dirac distribution function and normalizing factor $V^{-2}$ appears along with calculation of matrix element with Bloch functions of the band electrons.

Taking into account the identity for trace over spin variable $\sigma$, $\text{Tr}_\sigma (S_j s_e)(S_{j'} s_e) = \frac{1}{4} S_j S_{j'}$, and rearranging the **k** and **k'** in Eq. (11), one can obtain:

$$\hat{H}_{SS}^{(2)} = \frac{\beta^2}{4V^2} \sum_{k,k'(k'\neq k)} \frac{f(\varepsilon_k) - f(\varepsilon_{k'})}{\varepsilon_k - \varepsilon_{k'}} \sum_{j,j'(\neq j)} e^{i(k-k')(R_j - R_{j'})} S_j S_{j'}. \quad (12)$$

The procedure of second order perturbation theory implies to eliminate the points $k' = k$ from the sum in Eq. (12) [22]. Nevertheless, to restore analyticity of the integrant arising as a result of integral representation of the sums over **k'** and **k**, the expression (12) should be supplemented with the limit $k' \to k$ as it was stressed in original work of Yosida [8]. Applying to fraction in the Eq. (12), this limit produces $-\delta(\varepsilon_k - \varepsilon_F)$, where $\varepsilon_F$ is Fermi energy. Then the straightforward calculations yield the expression of the additive to Eq. (12) due to such treatment of $k \neq k'$ in sum (12) in the form:

$$\delta\hat{H}_{SS} = -\frac{\beta^2}{4V} D(\varepsilon_F) \sum_{j,j'(\neq j)} S_j S_{j'} = \frac{\beta^2 n_e \chi_e^{(1)}}{2V g_e^2 \mu_B^2} \sum_{j,j'(\neq j')} S_j S_{j'}. \quad (13)$$

where $D(\varepsilon_F) = 3n_e/4\varepsilon_F$ is the electron density of states of each spin branch at Fermi surface. Comparison of $\delta\hat{H}_{SS}$ with Eq. (8) shows that the integrant in integral approximation of the sums on **k** and **k'** in Eq. (12) transforms to smooth analytical function by adding the first order Hamiltonian $\hat{H}_{SS}^{(1)} = \delta\hat{H}_{SS}$ to the effective Hamiltonian represented spin-spin interaction in second order $\hat{H}_{SS}^{(2)}$. Calculation of these integrals in usual manner reproduces the common expression for RKKY interactions:

$$\hat{H}_{RKKY} = \hat{H}_{SS}^{(1)} + \hat{H}_{SS}^{(2)} = -\frac{1}{2} \sum_{j,j'(\neq j)} J_{RKKY}(R_{j,j'}) S_j S_{j'}, \quad (14)$$

where the effective constant of indirect interaction between the ions distanced on $R_{j,j'}$ is

$$J_{RKKY}(R_{j,j'}) = \frac{2\beta^2 k_F^3 \chi_{Pauli}}{\pi g_e^2 \mu_B^2} F_{RKKY}(2k_F R_{j,j'}) \quad (15)$$

and

$$F_{RKKY}(x) = \frac{x \cos x - \sin x}{x^4}. \quad (16)$$

The Eq.(14) along with definitions (15) and (16) is usually called "indirect RKKY exchange interaction". Each term $J_{RKKY}(R_{j,j'})S_j S_{j'}$ in sums (14) corresponds to a finite pair spin-spin interaction. For this reason, coefficient ½ guarantees each pair accounts once.

The expression (15) is obtained for degenerated electrons or holes and is not applicable for a single carrier case. However, Eq. (1a) reflects the dependence of CIEI exchange constant on the density of a carrier at a single LSM. The actual dependence of $k_F$ and $\chi_{Pauli}$ on carriers' concentration $n_e$ establishes the inverse proportionality of the prefactor of $F_{RKKY}(x)$ in (15) to carriers' density in power 4/3. Therefore, the enlargement of crystal volume $V$ under fixed number of carriers diminishes the $J_{RKKY}$ as $1/V^{4/3}$ that resembles Eq. (1a).

Let us emphases that the second order for indirect interaction in the form of Eq. (14) provides a finite strength of the effective exchange interaction (15) for any finite inter-ion distance $|R_{j,j'}|$ in case of finite carriers' concentration. From this standpoint, the addition (13) to the Hamiltonian (12) in the form of first order (8) does not modify this interaction in the limit V→∞ at $n_e$=const, because $\hat{H}_{SS}^{(1)} \xrightarrow[V\to\infty]{} 0$. Nevertheless, this conclusion fails if we turn from local (intensive) individual LSM magnetic property to thermodynamic (extensive) properties like a phase state in the entire system. As has been shown, thermodynamic limit does not vanish the molecular field stemmed from Hamiltonian $H_{SS}^{(1)}$ (5) provided a finite concentration $n_m$. Therefore, we should expect additive contributions to thermodynamic potentials stemmed from the first and second orders of Hamiltonians $\hat{H}_{SS}^{(1)}$ and $\hat{H}_{SS}^{(2)}$ [22]. In particular, exactly solvable model for spin cluster [12] verify this general result.

## 4. RKKY modelling of phase transition

Let us consider the critical temperature of FM phase transition calculated for RKKY interaction in Weiss field approximation. The latter imposes the equal mean values $\langle S_j \rangle = \langle S \rangle$ for all LSMs so that Weiss field

$$B_W = -\frac{1}{g_m \mu_B} \sum_{j'} J_{RKKY}(R_{j,j'}) \langle S_{j'} \rangle \qquad (17)$$

evenly polarizes each magnetic ion. Resulting spin polarization can be found in terms of Brillouin function $B_S$ as $\langle S \rangle = -S B_S(g_m \mu_B B_W S/T)$. Last equation evaluates the critical temperature of FM phase transition as

$$T_c = \frac{1}{3} S(S+1) \sum_{j'} J_{RKKY}(R_{j,j'}). \qquad (18)$$

Assuming a random LSM distribution with constant density $n_m$, the integral representation of the sum over $j'$ reads

$$\sum_{j'} J_{RKKY}(R_{j,j'}) \to n_m \int_0^\infty J_{RKKY}(r) 4\pi r^2 dr. \qquad (19)$$

Substitution this approximation into Eq. (18) reproduces the expression (6) for Curie temperature obtained in terms of first order perturbation theory. Obviously, applying another approximation for LSM distribution (such as lattice gas or correlated LSM dispensation) violates identity of the Eqs (18) and (6).

To clarify such coincident let us go back to the Eq. (11) and treat it as energy of spin $S_j$ in the Weiss effective field

$$B_W^{(2)} = \frac{\beta^2}{2 g_m \mu_B V^2} \sum_{k,k'(\neq k)} \frac{f(\varepsilon_k) - f(\varepsilon_{k'})}{\varepsilon_k - \varepsilon_{k'}} \sum_{j,j'(\neq j)} e^{i(k-k')(R_j - R_{j'})} \langle S_{j'} \rangle. \qquad (20)$$

The independence of $\langle S_j \rangle = \langle S \rangle$ on specific spin location immediately simplifies summation over $j'$. If ideal gas models the magnetic ion distribution over the crystal, the sum over $j'$ yields non-zero result only for $\boldsymbol{k} = \boldsymbol{k'}$ that, in turn, nullifies $B_w^{(2)}$. This result demonstrates that the conventional utilization of homogeneous distribution of the magnetic ions results in exclusion of second order perturbation. On the other hand, supplementing the singularity at $\boldsymbol{k} = \boldsymbol{k'}$ with the limit $\boldsymbol{k} \to \boldsymbol{k'}$ reproduces the $B_w^{(1)}$ (12) that is virtually included to $B_w^{(2)}$. It means that $B_w^{(1)}$ (12) does not appear once again as an independent contribution to Weiss field (19). Thus, a simple evaluation of $T_C$ in 1-th approximation (8) turn out equivalent to more complicated summation of spin-spin RKKY interactions over locations of magnetic ions in the crystal.

However, our approach can be applicable to more realistic LSM's distribution, which is not equivalent to constant probability independent on lattice structure. As first obvious generalization, let us consider a lattice gas approximation rather than free gas model. The simplest way to estimate the effect of lattice gas distribution consists in explicit limitation of the minimal inter-ion distance $d$. At larger distance, the constant probability, proportional to concentration $n_m$, still approximates the discrete distribution function over the lattice sites. Such elaboration diminishes the strength of the molecular field by the factor $\delta = \frac{4\pi}{3} n_m d^3$ provided $k_F d \ll 1$. This correction may be not too small compared with 1. For example, in DMS with zinc blende lattice the short-range space correlation modifies the expression (8) for Curie temperature by factor $1 - \frac{\sqrt{2}\pi}{3}x$ that reduces $T_C$ on 15% at LSM substitution level $n_m \Omega = x = 10\%$ ($\Omega$ is a volume of semiconductor primitive cell). Apparently, same result produces the RKKY approach [Eq. (14)] provided the integration over reducing domain $(d, \infty)$ takes into account the space correlations in Eq. (20).

Another, not trivial example discloses the case of approximate estimation of $T_C$ based on decreasing RKKY's interaction strength (15) for the remote LSMs. This diminishing interaction also results in decreasing the collective effect of the remote LSMs to the Weiss field. The latter can be evaluated using RKKY interaction of some LSM located in $\boldsymbol{R}_j$ site with a finite number LSMs surrounding $\boldsymbol{R}_j$ [10]. In such a case the contribution of first order (8), (13) vanishes assuming large crystal volume and a finite number $N_{m'}$ of the LSMs located at lattice sites $\boldsymbol{R}_{j'}$ inside an allotted volume $V'$ which maintains concentration $n_m = N_{m'} / V'$. Thus, the account of the finite number of LSMs might approximately evaluates the Weiss field in terms of the Eq. (20) derived in second order on CIEI. Appearance of non-zero effect of Eq. (20) looks not surprising since the summation over the limited numbers $\boldsymbol{R}_j$ in the finite volume V' admits a finite contributions of the wave vectors that lie beyond the specific electronic states with k=k'. Then, the estimation of $T_C$ assumes applying $B_w^{(2)}$ to all LSMs in whole crystal. Detail analysis of the Weiss field $B_w^{(2)}$ in a LSM cluster imbedded into large crystal is beyond the scope of current paper. Note however, that a cluster with $N_{m'}$ LSMs in a volume $V'$ mimics the actual DMS with fixed LSM concentration $n_m$, where the Weiss field is $B_w^{(1)}$. Such intuitive approach supposes that $B_w^{(2)}$ well approximates $B_w^{(1)}$ at large enough $V'$. Correctness of such approximation estimates a deviation of the integral in Eq. (19) from that calculated for the finite upper limit $2k_F(V')^{1/3}$. As seen, either consideration does not assume interference of $B_w^{(2)}$ and $B_w^{(1)}$ for Weiss field evaluation.

The $T_C$ accurate estimation supposes incorporating an inter-ion exchange interaction not related with band carriers to the formalism developed above. This antiferromagnetic (AFM)-type interaction inside of the nearest neighboring pairs of LSMs may exceed the effect of carrier-induced exchange field and thermal energy $T$ and establishes their total zero spin moments. As a result, such nearest neighboring LSMs drop out of consideration that appears as an effective LSM density x' < x [23]. Introducing the phenomenological parameter x' improves matching of experimental data and calculations for spin splitting of exciton spectra. Such approximation of AFM inter-ion interaction is also clearly applicable to consideration the magnetic (including FM) properties of DMS in terms of Weiss field calculated in first order. Nevertheless, the exclusion of AFM pairs from consideration enhance the deviation of actual LSM distribution from the ideal gas model along with consequent difference of the results obtained in second order approximation with ones a mean field threated in first order [3-5].

## 5. Conclusion

This paper clearly demonstrates that molecular field approximation applied to the second order expression of LSM indirect interaction in form of sum of exponentials (12) reveals the nulling of this contribution for all electronic states with wave vectors $\mathbf{k} \neq \mathbf{k}'$ provided free gas modelling of LSM distribution. The points $\mathbf{k} = \mathbf{k}'$ must be excluded in formalism of second ordeher perturbation theory. On default, however, these exclusive states commonly append to the formalism presuming the limit $\mathbf{k} \to \mathbf{k}'$ evaluates effect that does not dependent on magnetic ions locations. From this standpoint, the space averaging of the RKKY interaction (16) results in its homogeneous part of interaction, which appears as a supplement to indirect interaction in the limit $\mathbf{k} \to \mathbf{k}'$ coincided with the effective Hamiltonian (8) of first order interaction. In other words, the molecular field treatment of RKKY interaction produces the same result that can be obtained with taking into account only first order perturbation by CIEI. However, this is not a case for more advanced theories, which take into account a randomizing of the local exchange fields at particular LSM locations [14, 15]. Note also that utilizing a dispersionless band structure [13] violates smallness of the exchange constant $\beta N_0$ compared to bandwidth $W \to 0$.

In addition, the present paper appends the previous analyzes with particular second order effects stemmed from the difference of the lattice gas and ideal gas distributions for LSM in DMS.

## Acknowledgment

The participation of SMR in this work was partly supported by the project II-05-20. №11 of the Physics & Astronomy department of NAS Ukraine program for the support of the researches priority for the state.

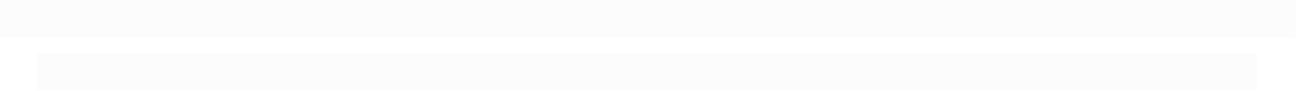